\documentclass{article}

\usepackage{amssymb}
\usepackage{amsmath}
\usepackage{latexsym}
\usepackage[dvips]{graphicx}

\def\abstract#1{\vskip 7mm 
        \begin{center}{\large Abstract}\par \smallskip
                \begin{minipage}[c]{12cm}
                        \small #1
                \end{minipage}
        \end{center}
}
\def\title#1{\begin{center}{\Large\bf #1}\end{center}}
\def\author#1{\vskip 5mm \begin{center}{#1}\end{center}}
\def\address#1{\begin{center}{\it #1}\end{center}}
\def\ead#1{\centerline{e-mail : #1}}

\begin{document}

\title{Black Hole Evaporation and Generalized 2nd Law with Nonequilibrium Thermodynamics
\footnote[2]{This report is a short summary of three papers~\cite{ref-sst.rad,ref-gsl.sst,ref-evapo.sst}. A full combined and rearranged version will be published as an invited contribution chapter (titled BH evaporation as a nonequilibrium process) in an edited book {\it Classical and Quantum Gravity Research Progress} (tentative title), Nova Science Publisher.}
\footnote[3]{Based on proceedings and talks given at APCTP Jeju Meeting on Gravitation and Cosmology (Gallery House, Jeju, Korea, 2007), Dynamics and Thermodynamics of Black Holes and Naked Singularities II (Politecnico di Milano, Milan, Italy, 2007) and 16th General Relativity and Gravitation (Niigata Prefectural Civic Center, Niigata, Japan, 2006)}}
\author{Hiromi Saida}
\address{Department of Physics, Daido Institute of Technology, Minami-ku, Nagoya 457-8530, Japan}
\ead{saida@daido-it.ac.jp}

\abstract{
In general, when a black hole evaporates, there arises a net energy flow from black hole into its outside environment due to Hawking radiation and energy accretion onto black hole. 
The existence of energy flow means that the thermodynamic state of the whole system, which consists of a black hole and its environment, is in a nonequilibrium state. 
To know the detail of evaporation process, the nonequilibrium effects of energy flow should be taken into account. 
The nonequilibrium nature of black hole evaporation is a challenging topic including issues of not only black hole physics but also nonequilibrium physics. 
Using the nonequilibrium thermodynamics which has been formulated recently, this report shows: 
(1) the self-gravitational effect of black hole which appears as its negative heat capacity guarantees the validity of generalized 2nd law without entropy production inside the outside environment, (2) the nonequilibrium effect of energy flow tends to shorten the evaporation time (life time) of black hole, and consequently specific nonequilibrium phenomena are suggested. 
Finally a future direction of this study is commented. 
This report summarizes three rather long papers~\cite{ref-sst.rad,ref-gsl.sst,ref-evapo.sst} without loading readers too much into the detail of discussions and analyses.
}

\section{Challenge to nonequilibrium nature of black hole evaporation}
\label{sec-challenge}

The black hole evaporation is one of interesting phenomena in black hole physics~\cite{ref-hr}. 
A direct treatment of time evolution of the evaporation process suffers from mathematical and conceptual difficulties; 
the mathematical one will be seen in the dynamical Einstein equation in which the source of gravity may be a quantum expectation value of stress-energy tensor of Hawking radiation, and the conceptual one will be seen in the definition of dynamical black hole horizon. 
Therefore an approach based on the black hole thermodynamics~\cite{ref-hr,ref-bht,ref-gsl} is useful.

Exactly speaking, dynamical evolution of any system is a nonequilibrium process. 
If and only if thermodynamic state of the system under consideration passes near equilibrium states during its evolution, its dynamics can be treated by an approximate method, the so-called {\it quasi-static process}. 
In this approximation, it is assumed that the thermodynamic state of the system evolves on a path lying in the state space which consists of only thermal equilibrium states, and the time evolution is described by a succession of different equilibrium states. 
However once the system comes far from equilibrium, the quasi-static approximation breaks down. 
In that case a nonequilibrium thermodynamic approach is necessary. 
For dissipative systems, the heat flow inside the system can quantify the degree of nonequilibrium nature~\cite{ref-eit,ref-sst,ref-anti.sst}.

For the black hole evaporation, when its horizon scale is larger than Planck size, it is relevant to describe the black hole itself by equilibrium solutions of Einstein equation, Schwarzschild, Reissner-Nortstr\"{o}m and Kerr black holes, because the evaporation proceeds extremely slowly and those equilibrium solutions are stable under gravitational perturbations~\cite{ref-chandra}. 
The slow evolution is understandable by the Hawking temperature~\cite{ref-hr} which is regarded as an equilibrium temperature of black hole,
\begin{equation}
 T_g := \dfrac{m_{pl}^2}{8 \pi M} \, ,
\label{eq-Tg}
\end{equation}
where $M$ is the black hole mass, $m_{pl}$ is the Planck mass and the units $c = \hbar = k_B = 1$ and $G = 1/m_{pl}^2$ are used. 
Obviously a classical size black hole ($M \gg m_{pl}$) has a very low temperature. 
This means a very weak energy emission rate by the Hawking radiation which is proportional to $(2 G M)^2\,T_g^4$ due to the Stefan-Boltzmann law. 
Therefore the quasi-static approximation works well for the black hole itself during its evaporation process. 
However the outside environment around black hole may not be described by the quasi-static approximation because of the energy flow due to the Hawking radiation. 
The Hawking radiation causes an energy flow in the outside environment, and that energy flow drives the outside environment out of equilibrium. 
As indicated by eq.\eqref{eq-Tg}, the black hole temperature and the energy emission rate by black hole increase as $M$ decreases along the evaporation. 
The stronger the energy emission, the more distant from equilibrium the outside environment. 
Therefore the nonequilibrium nature of the outside environment becomes stronger as the black hole evaporation proceeds. 
At the same time, the quasi-static approximation is applicable to the black hole itself since equilibrium black hole solutions are stable under gravitational perturbation. 
Hence, in studying detail of evaporation process, while the black hole itself is described by quasi-static approximation, but the nonequilibrium effects of the energy flow in the outside environment should be taken into account.

In the above paragraph, the energy accretion onto black hole is ignored. 
However if the temperature of outside environment is non-zero and lower enough than the black hole temperature, then the black hole evaporates under the effect of energy exchange due to the Hawking radiation and the energy accretion. 
In this case the same consideration explained in the above holds and we recognize the importance of the net energy flow from black hole to outside environment. 
Dynamical behaviors of black hole evaporation will be described well by taking nonequilibrium nature of the net energy flow into account.

In next section, we start with constructing a nonequilibrium thermodynamics suitable to the nonequilibrium nature of black hole evaporation~\cite{ref-sst.rad}. 
Then it is applied to black hole evaporation. 
Section~\ref{sec-gsl} summarizes~\cite{ref-gsl.sst}, which reveals that the generalized second law is guaranteed not by self-interactions of matter fields around black hole which cause self-production of entropy inside the matters, but by the self-gravitational effect of black hole appearing as its negative heat capacity in eq.\eqref{eq-capacity}. 
Section~\ref{sec-evapo} summarizes~\cite{ref-evapo.sst}, which reveals that the nonequilibrium effect tends to accelerate the evaporation process and, consequently, gives a suggestion about the end state of quantum size black hole evaporation in the context of the information loss paradox. 
Finally section~\ref{sec-future} concludes this report with comments for future direction of this study.

Throughout this report except for eq.\eqref{eq-Tg}, Planck units are used, $c = \hbar = G = k_B = 1$.

\section{Basic consideration and Steady state thermodynamics}
\label{sec-sst}

According to the black hole thermodynamics~\cite{ref-bht}, a stationary black hole can be regarded as a {\it black body}, an object whose thermodynamic state is in thermal equilibrium. 
For simplicity, let us consider a Schwarzschild black hole. Its equations of states as a black body are
\begin{equation}
 E_g = \frac{1}{8 \pi T_g} = \frac{R_g}{2} \quad , \quad
 S_g = \frac{1}{16 \pi T_g^2} = \pi R_g^2 \, ,
\label{eq-eos}
\end{equation}
where $E_g$ is mass energy, $R_g$ is areal radius of horizon, $T_g$ is the Hawking temperature, and $S_g$ is the Bekenstein-Hawking entropy. 
It is obvious that the heat capacity of a black hole is negative, 
\begin{equation}
 C_g := \frac{dE_g}{dT_g} = - \frac{1}{8 \pi T_g^2} = - 2 \pi R_g^2 < 0 \, .
\label{eq-capacity}
\end{equation}
The negative heat capacity is a peculiar property to self-gravitating systems~\cite{ref-sgs}. 
Therefore the energy $E_g$ encodes well the self-gravitational effects of black hole on its own thermodynamic state.
Furthermore it has already been revealed that, using the Euclidean path-integral method for a black hole spacetime and matter fields on it, an equilibrium entropy of whole gravitational field on a black hole spacetime is given by the Bekenstein-Hawking entropy~\cite{ref-entropy}. 
This means $S_g$ in eq.\eqref{eq-eos} is the equilibrium entropy of whole gravitational field on black hole spacetime, and the gravitational entropy vanishes if there is no black hole horizon. 
Hence we find that energetic and entropic properties of a black hole are encoded in the equations of states~\eqref{eq-eos}.

\begin{figure}[t]
 \begin{center}
 \includegraphics[height=40mm]{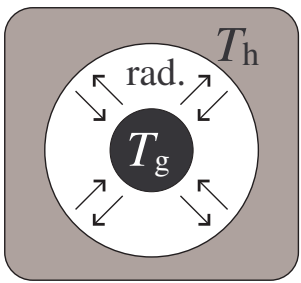}
 \end{center}
\caption{NE model. The radiation fields sandwiched by black bodies are in the two-temperature steady state.}
\label{fig-1}
\end{figure}

As mentioned section~\ref{sec-challenge}, the nonequilibrium nature arises in the matter fields around black hole due to energy flows by Hawking radiation and by matter accretion from the outside environment onto the black hole. 
In general, nonequilibrium phenomena is one of the most difficult subjects in physics, and it is impossible at present to treat the nonequilibrium nature of black hole evaporation in a full general relativistic framework. 
For simplicity, let us assume that the matter fields around black hole are not only massless but also non-self-interacting (gas of collisionless particles), and call the matter fields {\it the radiation fields}. 
Furthermore, we resort to a simplified model which reflects the nonequilibrium effects of energy exchange between a black hole and its environment~\cite{ref-evapo.sst}:
\begin{description}
\item[Nonequilibrium Evaporation (NE) model:] Put a spherical black body of temperature $T_g$ in a heat bath of temperature $T_h (< T_g)$, where the equations of states of the spherical black body are eq.\eqref{eq-eos} and we call the black body {\it the black hole}. 
Let the heat bath (the outer black body of temperature $T_h$) be made of ordinary materials of positive heat capacity. 
Then, hollow a spherical region out of the heat bath around the black hole as seen in fig.\ref{fig-1}. 
The hollow region is a shell-like region which is concentric with the black hole and separates the black hole and the heat bath, and filled with matter fields emitted by them. 
For simplicity the matter fields are non-self-interacting massless matter fields, which we call {\it the radiation fields}. 
Furthermore, in order to validate using the ``equilibrium (static)'' equations of states~\eqref{eq-eos}, we assume the following:
\item[Quasi-equilibrium assumption:] Time evolution in the NE model is not so fast that the thermodynamic states of black hole and heat bath at each moment of their evolution are approximated well by equilibrium states individually. 
This means thermodynamic state of black hole evolves along a sequence of equilibrium states in the state space during the time evolution of the whole system. 
The same is true of the heat bath. (Recall a quasi-static process in ordinary thermodynamics). 
Then it is valid to use eq.\eqref{eq-eos} which describes a static (Schwarzschild) black hole. 
Also, since Schwarzschild black hole is not a quantum one, it is required,
\begin{equation}
 R_g > 1 \, .
\label{eq-qe}
\end{equation}
\end{description}
Here note that the temperature difference ($T_g > T_h$) causes a net energy flow from the black hole to the heat bath (a relaxation process), and $T_g$ increases along the decrease of $E_g$ due to the negative heat capacity given in eq.\eqref{eq-capacity}. 
This relaxation process describes the black hole evaporation in the framework of NE model.

Here let us comment about terminology. 
There may be an objection that the term ``relaxation'' is not suitable for the case of increasing temperature difference. 
But in this report, please understand it means the time evolution arising in isolated inhomogeneous systems.

As mentioned above, we consider the radiation fields in the hollow region. 
When the number of independent helicities in radiation fields is $N$, the Stefan-Boltzmann constant becomes $\sigma = N \pi^2/120$ ($N=2$ for photon). 
Note that, due to eq.\eqref{eq-qe} of quasi-equilibrium assumption, the NE model describes a semi-classical stage of evaporation, $T_g < 1$. Therefore it is appropriate to evaluate $N$ by the number of independent states of standard particles (quarks, leptons and gauge particles of four fundamental interactions), $N = O(100)$.

If a full general relativistic treatment is possible, the Hawking radiation experiences the curvature scattering to form a spacetime region filled with interacting matters. 
Then some fraction of Hawking radiation is radiated back to the black hole from that region. 
The heat bath in the NE model is understood as a simple representation of not only matters like accretion disk but also such region formed by curvature scattering. 
However the gravitational redshift on radiation fields in the hollow region (the so-called {\it grey body factor}) is ignored in the NE model.

In the hollow region, the nonequilibrium state of radiation fields (at each moment of its time evolution) are described by a simple superposition of a state of temperature $T_g$ and that of $T_h$, since the radiation fields are collisionless gas. 
That is, the radiation fields are in a two-temperature nonequilibrium state at each moment of time evolution. 
Because of the quasi-equilibrium assumption, the two-temperature nonequilibrium state of radiation fields (at each moment of time evolution) can be described well by a nonequilibrium state which possesses a net stationary energy flow from the black hole to the heat bath. 
We call this macroscopically stationary nonequilibrium state {\it the steady state}. 
Then the time evolution of radiation fields is expressed by a {\it quasi-steady process} in which the thermodynamic state of radiation fields evolves on a path lying in the state space which consists of steady states, and the time evolution is described by a succession of different steady states. 
Hence, we need a thermodynamic formalism of two-temperature steady states for radiation fields in order to analyze the black hole evaporation process in the framework of NE model.

The steady state thermodynamics (SST) has already been formulated, and the 0th, 1st, 2nd and 3rd laws are established in~\cite{ref-sst.rad}. 
By the SST for radiation fields, the steady state energy $E_{rad}$ and steady state entropy $S_{rad}$ of the radiation fields are given as
\begin{equation}
 E_{rad} = \int dx^3\, 4 \sigma
           \left(\, g_g(\vec{x})\, T_g^4 + g_h(\vec{x})\, T_h^4\, \right) \quad , \quad
 S_{rad} = \int dx^3\, \frac{16 \sigma}{3}
           \left(\, g_g(\vec{x})\, T_g^3 + g_h(\vec{x})\, T_h^3\, \right) \, ,
\label{eq-sst}
\end{equation}
where $\vec{x}$ is a spatial point in the hollow region, $g_g(\vec{x})$ is given by the solid-angle (divided by $4 \pi$) covered by the black hole seen from $\vec{x}$, and $g_h(\vec{x})$ is that given by the heat bath. By definition, $g_g(\vec{x}) + g_h(\vec{x}) \equiv 1$.

Although the NE model may be too simple, let us try to investigate nonequilibrium effects of the net energy flow from black hole to its outside environment in the framework of NE model.

Before proceeding to review of black hole evaporation, let us comment about the SST for radiation fields. 
Apart from black hole physics, in the existing theories of nonequilibrium dissipative systems, a heat flow plays the role of well-defined nonequilibrium order parameter, a state variable quantifying the degree of nonequilibrium nature of the dissipative system. 
On the other hand, because of the non-dissipative nature of radiation fields, it has been noticed that a heat bath becomes ill-defined as a nonequilibrium order parameter of nonequilibrium radiation fields~\cite{ref-noneq.rad}. 
One of the reason is that the ``heat'' flow arises due to dissipation and no heat flow exists in non-dissipative systems like the radiation fields. 
However a well-defined nonequilibrium order parameter instead of heat flow has not been found. 
At least for steady states (not for general nonequilibrium states), it is proposed in~\cite{ref-sst.rad} that a well-defined nonequilibrium order parameter of radiation fields is the temperature difference (e.g. $T_g - T_h$ for the NE model). 
This is one of peculiar properties of radiation fields. 
There are some other peculiar thermodynamic properties of radiation fields due to the non-dissipative nature, which are discussed in~\cite{ref-sst.rad}. 
Although a full understanding of SST for radiation fields is not necessary for black hole evaporation at present, it may be useful to understand radiative phenomena in astrophysics.

\section{Physical essence of generalized 2nd law}
\label{sec-gsl}

For simplicity, consider a black hole evaporation in an empty space. 
The black hole entropy $S_g$ decreases as the mass energy $E_g$ decreases along the evaporation process. 
Then the generalized second law (GSL) {\it conjectures} that the total entropy $S_{tot} = S_g + S_m$ increases ($dS_{tot} > 0$) as the evaporation process proceeds, where $S_m$ is the entropy of matter fields of Hawking radiation. 
There are three candidates for the physical origin of GSL:
\begin{description}
\item[Origin of GSL (a):] Self-interactions of matter fields of Hawking radiation (collision of particles and self-gravitational interaction).
\item[Origin of GSL (b):] Gravitational interaction between black hole and matter fields of Hawking radiation (curvature scattering, lens effect, gravitational redshift and so on).
\item[Origin of GSL (c):] Increase of temperature $T_g$ along the evaporation process due to the negative heat capacity $C_g < 0$ (the evaporation denotes $dE_g < 0 \,\, \Rightarrow \,\, dT_g = dE_g/C_g > 0$). In other words, this candidate~(c) is the self-gravitational effect of black hole, since negativity of heat capacity is due to the self-gravitation of the system under consideration.
\end{description}
In the existing proofs of GSL~\cite{ref-gsl.proof}, all of these candidates are considered and it has remained unclear which of these dominates over the others. 
If the GSL would be proven by considering a situation keeping one of them and discarding the others, then we can conclude that the one kept is the essence of GSL. 
In reference~\cite{ref-gsl.sst}, it has been revealed that the third candidate~(c) is the essence of GSL. 
The outline of discussion in~\cite{ref-gsl.sst} is summarized as follows.

Let us consider about the candidates~(a) and~(b). 
The~(a), self-interaction of matter fields of Hawking radiation, denotes clearly the positive entropy production rate inside the matter fields. 
This means that the self-relaxation (self-production of entropy) of the matter fields occurs due to the self-interactions. 
The~(b), gravitational interaction between the black hole and the matter fields, causes the relaxation of the matter fields as well and the matter entropy is produced. 
Therefore~(a) and~(b) give a positive entropy production rate inside the matter fields of Hawking radiation. 
Matter entropy increases by~(a) and~(b) during propagating in the outside space of black hole. 
On the other hand, the candidate~(c), self-gravitational effect of black hole, does not cause the entropy production inside the matter fields. 
The~(c) denotes that the entropy of the matter fields just at the moment of its emission at the black hole horizon (the ``inherent'' entropy of matter fields) increases as the black hole evaporation process proceeds, because the temperature $T_g$ of the ``entropy source'' increases. 
This is not the self-production of entropy by matter fields, but the emission of entropy by the black hole.

\begin{figure}[t]
 \begin{center}
 \includegraphics[height=80mm]{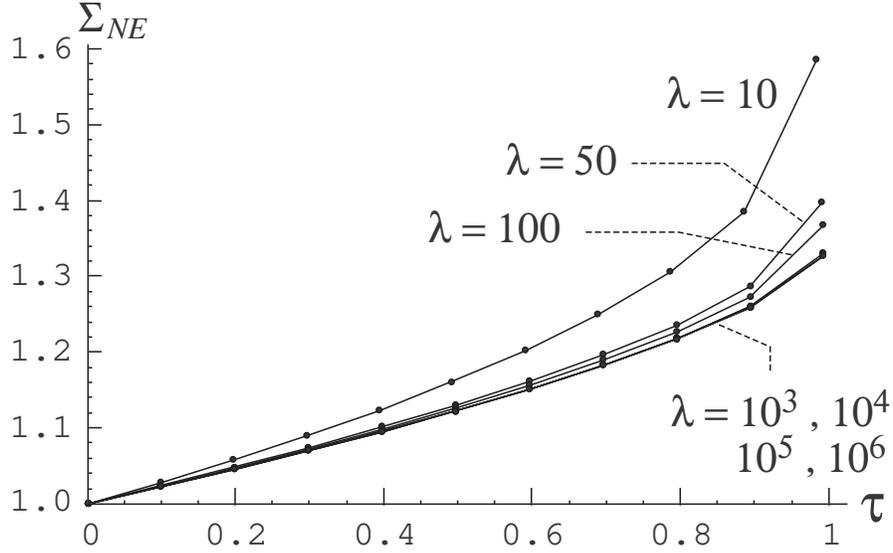}
 \end{center}
\caption{Time evolution of normalized total entropy $\Sigma_{NE}$ in the framework of NE model.}
\label{fig-2}
\end{figure}

The NE model includes~(c) due to the equations of state~\eqref{eq-eos}, but not~(a) because the radiation fields are of massless non-self-interacting. 
Here recall that~(b) causes a relaxation of matter fields and, as mentioned section~\ref{sec-sst}, the heat bath in NE model is understood as a very simple and rough model representing~(b). 
Then a black hole evaporation which reflects only~(c) can be obtained by removing the heat bath from the NE model and makes the radiation fields spread out into an infinitely large flat spacetime (black hole evaporation in an empty space with ignoring the grey body factor). 
With removing the heat bath, the total entropy $S_{tot} := S_g + S_{rad}$ is calculated concretely using $S_g$ in eq.\eqref{eq-eos} and $S_{rad}$ in eq.\eqref{eq-sst}. 
Time evolution of $S_{tot}(t)$ is induced by the energy emission by black hole which is estimated by the Stefan-Boltzmann law $dE_g/dt = - \sigma\,T_g^4\,A_g$, where $A_g = 4\pi\,R_g^2$ is surface area of black hole and a time $t$ corresponds to a proper time of rest observer distant from black hole. 
Since $S_{tot}(t)$ has too complicated functional form of $t$, analytic proof of $dS_{tot}/dt > 0$ will be difficult.
Then we plot $S_{tot}(t)$ numerically. 
The result is shown in fig.\ref{fig-2}. 
In this figure, $\tau := t/t_{empty}$ is a time normalized by the evaporation time (life time) of black hole evaluated by the Stefan-Boltzmann law, $\Sigma_{NE} := S_{tot}(\tau)/S_{tot}(0)$ is a total entropy normalized by its initial value. 
Furthermore $\lambda$ is defined as $\lambda := t_{empty}/R_g(0)$, where the initial radius of black hole $R_g(0)$ gives a typical time scale of radiation fields to spread out into infinitely large space. 
Using the Stefan-Boltzmann law we can find $\lambda \propto R_g(0)^2$. 
The larger the parameter $\lambda$, the slower the evaporation proceeds and the more valid the quasi-equilibrium assumption. 
As shown in fig.\ref{fig-2}, $dS_{tot} > 0$ is found and it is concluded that~(c) is the physical essence to guarantee GSL:
\begin{description}
\item[Conclusion I by~\cite{ref-gsl.sst}:] It is not the interactions of matter fields around black hole, but the self-gravitation of the black hole (negative heat capacity) which guarantees the validity of GSL.
\end{description}

Furthermore, fig.\ref{fig-2} shows that the final value of total entropy is about $\Sigma_{NE}(t_{empty}) \sim 1.33$ for sufficiently large $\lambda$. 
On the other hand, if we consider a black hole evaporation including the candidates~(a) and~(d) together with~(c), then the total entropy of such case will become larger than $\Sigma_{NE}$ which is evaluated only with~(c). 
Therefore $\Sigma_{NE}$ in fig.\ref{fig-2} expresses a lower bound of total entropy of any possible black hole evaporation process including~(a), (b) and~(c). 
Hence we can find another conclusion:
\begin{description}
\item[Conclusion II by~\cite{ref-gsl.sst}:] For black hole evaporation in a full general relativistic framework including self-interaction of matter fields and gravitational interaction between black hole and matter fields, the final value of total entropy should be larger than $1.33 \times \text{its initial value}$.
\end{description}
Due to eq.\eqref{eq-qe} of quasi-equilibrium assumption, the NE model is valid only for semi-classical evaporation stage. 
However we expect the conclusion II holds even for the quantum evaporation stage, because the entropy should increase along any classical and quantum evolution process of isolated systems.

Turn our discussion to a possible objection. 
Concerning the entropy of self-gravitating systems like black holes, stars and so on, one may think as follows: 
When an inter-stellar gas collapses to form a star, the self-gravitational effect of that gas decreases its entropy. 
Then the black hole evaporation under the self-gravitational effect of black hole may not result in an increase of total entropy. 
Let us try to answer to this objection.

When an interstellar gas collapses to form a star, it is commonly believed that there arises the increase of net entropy of total system which consists of the collapsing gas and the radiated matters from the gas. 
However the self-gravitational effect of the collapsing gas causes the decrease of entropy of the collapsing gas. 
It is briefly explained as follows: 
Since the pressure of that gas at its surface is zero, the ``loss'' of energy of collapsing gas par a unit time $\Delta E$ due to energy emission is actually the loss of heat due to the first law of thermodynamics,
\begin{equation}
 \Delta E \sim T \, \Delta S \, 
\label{eq-gsl.sum.1}
\end{equation}
where $T$ is the temperature of collapsing gas, and $\Delta S$ is the ``loss'' of entropy of collapsing gas par a unit time. 
The energy loss $\Delta E$ is the minus of the luminosity $L$ of collapsing gas, $L = - \Delta E$. 
Then, with the assumption of local mechanical and thermal equilibrium of collapsing gas at each moment of its collapse, relation~\eqref{eq-gsl.sum.1} is rewritten more exactly to the following form~\cite{ref-sgs},
\begin{equation}
 L = - \frac{d [ \, U + \Omega \,]}{dt} \sim - T \Delta S \, ,
\end{equation}
where $U$ is the total internal energy of collapsing gas, and $\Omega$ is the total self-gravitational potential given by
\begin{equation}
 \Omega = \int_0^M \, dm \,\left( - \frac{G\, m}{r(m)} \, \right) \, ,
\end{equation}
where $G$ is Newton's constant, $M$ is mass of collapsing gas, and the radial distance from the center of gas $r(m)$ is expressed as a function of mass $m$ inside a sphere of radius $r$. 
From the above, we recognize that the radius $r(m)$ becomes smaller as the gas collapses, then the self-gravitational potential $\Omega$ decreases to result in the radiation of energy $L >0$ and the loss of entropy $\Delta S < 0$. 
This is a similar phenomenon to the evaporation of black hole itself with decreasing its entropy $dS_g < 0$. 
On the other hand, if we consider not a collapsing but an expanding self-gravitating gas, the radius $r(m)$ increases to result in the increase of entropy $\Delta S >0$. 
Here we point out that the Hawking radiation in black hole evaporation corresponds not to a collapsing gas but to an expanding gas. 
Therefore, if we consider the radiation fields of Hawking radiation including their self-gravitational effect, the entropy of radiation will increase during spreading out into an infinitely large space. 
This is just the origin~(a) of GSL. 
Hence the entropy of self-gravitating matter fields of Hawking radiation $S_m$ is larger than that of a non-self-gravitating Hawking field $S_{rad}$. 
This relation $S_m > S_{rad}$ together with the result $d(S_g + S_{rad}) > 0$ shown in fig.\ref{fig-2} imply the objection mentioned above is not true of the black hole evaporation process.

Finally let us recall a statement given in the beginning of previous paragraph; it is commonly ``believed'' that there arises the increase of net entropy of total system which consists of the collapsing gas and the radiated matters from the gas. 
One of the reasons why it is not proven but ``believed'' is that there has not been nonequilibrium thermodynamics to treat the net entropy. 
Although we considered the black hole evaporation in this section, a similar method based on the SST will be applicable to a star formation process including radiations from collapsing gas. 
Then ``believed'' will become ``proven''.

\section{Black hole evaporation interacting with outside environment}
\label{sec-evapo}

In general, a black hole evaporation proceeds under the effects of interactions with the outside environment around black hole (e.g matter accretion, curvature scattering and so on). 
The heat bath in NE model can be considered as a simple model of such interactions, and it is expected that the analysis of time evolution of NE model gives a thermodynamic understanding of black hole evaporation interacting with outside environment.

Hereafter assume that the whole system of NE model (black hole, heat bath and radiation fields) is isolated from the outside world around heat bath (micro-canonical ensemble). 
It is useful to divide the whole system into two sub-systems X and Y. 
The X consists of the black hole and the ``out-going'' radiation fields emitted by black hole, which share the temperature $T_g$. 
The Y consists of the heat bath and the ``in-going'' radiation fields emitted by heat bath, which share the temperature $T_h$. 
Then the energy transport between X and Y is given by the Stefan-Boltzmann law,
\begin{equation}
 \frac{dE_X}{dt} = -\sigma \left( T_g^4 - T_h^4 \right) \, A_g \quad , \quad
 \frac{dE_Y}{dt} = \sigma \left( T_g^4 - T_h^4 \right) \, A_g \, ,
\label{eq-transport}
\end{equation}
where $A_g := 4 \pi R_g^2$, $E_X := E_g + E_{rad}^{(out)}$ ($E_{rad}^{(out)} := \int dx^3 4 \sigma g_g T_g^4$) is the energy of X, $E_Y := E_h + E_{rad}^{(in)}$ ($E_{rad}^{(in)} := \int dx^3 4 \sigma g_h T_h^4$, \,$E_h(T_h) =$ energy of heat bath) is the energy of Y, and $E_{rad}^{(out)} + E_{rad}^{(in)} = E_{rad}$ by definition (see eq.\eqref{eq-sst}). 
This energy transport is consistent with the isolated condition of the whole system, $E_{tot} := E_X + E_Y = constant$. 
In order to analyze these nonlinear equations~\eqref{eq-transport}, it is useful to consider the energy emission rate (luminosity) by the black hole, 
\begin{equation}
 J_{ne} := -\frac{dE_g}{dt} \, .
\end{equation}
The larger the value of $J_{ne}$, the more rapidly the energy $E_g$ of black hole decreases along its evaporation process. 
The stronger emission rate $J_{ne}$ denotes the acceleration of the evaporation process. 
Furthermore we compare $J_{ne}$ with 
\begin{equation}
 J_{empty} := \sigma T_g^4 A_g \, .
\end{equation}
This $J_{empty}$ is the energy emission rate by a black hole of the same mass $E_g$ evaporating in an empty space (as considered in previous section). 
Then, applying eq.\eqref{eq-transport} to $J_{ne}$, we obtain a relation,
\begin{equation}
 \frac{J_{ne}}{J_{empty}} =
 \left|\frac{C_g}{C_X} \right|\, \left( 1 - \frac{T_h^4}{T_g^4} \right) \, ,
\end{equation}
where $C_X = dE_X/dT_g$ is the heat capacity of sub-system X (see~\cite{ref-evapo.sst} for derivation). 
By a careful analysis of $J_{ne}/J_{empty}$ with a support of numerical solutions of eq.\eqref{eq-transport}, reference~\cite{ref-evapo.sst} gives a non-trivial result (where $t_{ne}$ denotes the time scale of evaporation in the framework of NE model, and $t_{empyt}$ denotes that in an empty space):
\begin{description}
\item[Conclusion by~\cite{ref-evapo.sst}:] When the thickness of the hollow region in the NE model (see figure~\ref{fig-1}) is thin enough, then $J_{ne}/J_{empty} < 1$ holds along the black hole evaporation in NE model and $t_{ne} > t_{empty}$ is obtained. 
When the hollow region is thick enough, then $J_{ne}/J_{empty} > 1$ holds along the black hole evaporation in NE model and $t_{ne} < t_{empty}$ is obtained. 
(See the following for the non-triviality of this result.)
\end{description}
One may naively think that the existence of the heat bath (``in-coming'' energy flow into the black hole) always decelerates the evaporation and $t_{ne} < t_{empty}$ does not occur. 
However the above result states that this naive sense does not work with a thick hollow region. 
In order to understand the failure of naive sense, it is important to recognize that the evaporation in an empty space (see previous section) is regarded as a relaxation process of ``isolated'' sub-system X keeping $E_X = constant$ during the evaporation process, however the evaporation in the NE model has energy extraction from X by Y due to the temperature difference $T_h < T_g$.

Then, there are two points to understand the failure of naive sense due to the energy extraction: 
The first point is that, because the energy extraction ($dE_X < 0$) occurs along the black hole evaporation ($dT_g > 0$) and the whole system is isolated ($dE_X + dE_Y = 0$), then the heat capacity of X is always negative $C_X := dE_X/dT_g = C_g + C_{rad}^{(out)} < 0$ in the framework of NE model. 
Here the heat capacity $C_{rad}^{(out)}$ of out-going radiation fields is positive since the radiation fields is an ordinary matter possessing positive specific heat. 
Note that, the larger the volume of the hollow region in NE model, the larger the heat capacity $C_{rad}^{(out)}$ and the smaller the absolute value $\left|C_X\right|$. 
This implies that, the more thick the hollow region, the more accelerated the increase of $T_g$ due to the relation $dT_g = \left|dE_X/C_X\right|$. 
Therefore, as the evaporation process proceeds, the energy extraction from X by Y (the increase of $T_g$) comes to dominate over the in-coming energy flow into the black hole. 
This means that the energy emission rate $J_{ne}$ is enhanced by the energy extraction from X by Y, and then it results in $t_{ne} < t_{empty}$ for the NE model of thick hollow region.

The second point to understand the failure of naive sense is that the energy transport equation $dE_g/dt = - J_{empty}$ of the evaporation in an empty space can not be obtained from those of NE model~\eqref{eq-transport}. 
One may expect that the empty case would be recovered by a limit operation, $T_h \to 0$ and $V_{rad} \to \infty$, where $V_{rad}$ is the volume of the hollow region. 
However this operation gives $R_g = \infty$ which is unphysical (see~\cite{ref-evapo.sst} for calculation). 
Therefore, the evaporation in an empty space can not be described as some limit situation of NE model. 
Hence, the naive sense which is based on a limit operation of NE model leads a mistake. 
Furthermore, if the empty case was a limit situation of NE model with infinite volume of hollow region, the evaporation time of the empty case would be zero ($t_{empty} \to 0$) according to the above conclusion. 
However we know $t_{empty} \neq 0$, and it also implies that the empty case is not a limit situation of NE model.

Next turn our discussion to the suggestion by the NE model. 
Because of eq.\eqref{eq-qe} in the quasi-equilibrium assumption, the NE model is valid only for semi-classical evaporation stage. 
The SST for the radiation fields enables us to estimate the value of quantities like $J_{ne}$, $E_g$ and $S_g$ at the end of semi-classical stage (at the onset of quantum evaporation stage) more precisely than the method based on ordinary thermodynamics. 
And it is found that $J_{ne}$ is much larger than $J_{empty}$:
\begin{description}
\item[Suggestion I by~\cite{ref-evapo.sst}:] Evaluation of $J_{ne}/J_{empty}$ implies that the black hole evaporation in the framework of NE model will end with a huge energy burst $J_{ne}$ stronger than that $J_{empty}$ of the evaporation in an empty space.
\end{description}

After the semi-classical evaporation stage, a quantum size black hole evaporation stage follows. 
We can guess physically what happens along the quantum evaporation stage using the NE model without referring to present incomplete theories of quantum gravity: 
If we assume that the black hole evaporates out completely at the end of quantum evaporation stage, then the whole system after the complete evaporation consists of heat bath and radiation fields without including quantum size black hole. 
Using the SST for radiation fields, we can calculate the total entropy of that whole system just after the complete evaporation of black hole. 
Also the total entropy of NE model just before the onset of quantum evaporation stage is calculated using the SST. 
If the complete evaporation of black hole is true of the case, the total entropy should increase due to the 2nd law of thermodynamics. 
However, a precise analysis done in~\cite{ref-evapo.sst} shows the total entropy decreases if we assume the complete evaporation of black hole. 
Hence, by the reductive absurdity, the followings are suggested:
\begin{description}
\item[Suggestion II by~\cite{ref-evapo.sst}:] A remnant of Planck size may remain at the end of the quantum evaporation stage in order to guarantee the increase of total entropy along the whole process of evaporation. 
\end{description}

This suggestion may also relate to the so-called {\it information loss paradox}~\cite{ref-info}. 
The information loss paradox points out an inconsistency between unitarity of quantum system and complete evaporation of black hole as follows: 
If a black hole evaporates out completely, then the thermal spectrum of Hawking radiation implies a thermal radiation remains after the complete evaporation. This also implies that, even if the initial state of black hole formation (gravitational collapse) consists of pure quantum state, but the state is transformed to a thermal state at the end of black hole evaporation. 
Such a transformation is not unitary. 
A some kind of information of initial state will be lost during the black hole evaporation process if it evaporates out completely.

However our suggestion II can propose a possibility that the remnant preserves the initial information, although we can not say anything about the mechanism how to preserve it. 
If it is true of the case, the information loss paradox disappears due to the nonequilibrium effect of black hole evaporation.

\section{Concluding comments}
\label{sec-future}

The NE model is not a full general relativistic model, and ignoring gravitational redshift and curvature scattering on radiation fields propagating in hollow region. 
Towards a general relativistic NE model, we have to extend the steady state thermodynamics to its general relativistic version. 
When we will construct a general relativistic steady state thermodynamics for radiation fields, the gravitational effects on radiation fields will be included in the NE model.

All of the results in this report are obtained by NE model and based on the steady state thermodynamics for radiation fields. 
Nonequilibrium thermodynamic approach may be a powerful tool for investigating the nonlinear and dynamical phenomena. 
While we considered the radiation fields which is of non-self-interacting, however nonequilibrium thermodynamics for ordinary dissipative systems has already been established in, for example, the extended irreversible thermodynamics~\cite{ref-eit}. 
It is applicable not to any highly nonequilibrium state but to state whose entropy flux is well approximated up to second order in the expansion by the heat flux of a nonequilibrium state under consideration. 
It is interesting to consider self-interacting matter field for the Hawking radiation and energy accretion, and apply the extended irreversible thermodynamics to those fields. 
Then we may find a variety of black hole evaporation phenomena. 
Furthermore apart from black hole evaporation, since accretion disks around black hole seem to consist of dissipative matters in realistic settings, the extended irreversible thermodynamics may give a unique new approach to investigate black hole astrophysics. 
And by applying the extended irreversible thermodynamics to a collapsing matter, one may give a progress in the research of gravitational collapse.

Apart from black hole physics, the steady state thermodynamics for radiation fields may be helpful to understand, for example, the free streaming in the universe like cosmic microwave background and/or the radiative energy transfer inside a star and among stellar objects. 
Also an example of possible application to a star formation process is explained at the end of section~\ref{sec-gsl}.



\begin{thebibliography}{99}
%
\bibitem{ref-hr}
 S.W.Hawking, {\it Commun.Math.Phys.} {\bf 43} (1975) 199
%
\bibitem{ref-bht}
 S.W.Hawking, {\it Phys.Rev.Lett.} {\bf 26} (1971) 1344 \\
 S.W.Hawking, {\it Phys.Rev.} {\bf D13} (1976) 191 \\
 J.M.Bardeen, B.Carter and S.W.Hawking, {\it Commun.Math.Phys.} {\bf 31} (1973) 161 \\
 W.Israel, {\it Phys.Rev.Lett.} { \bf 57} (1986) 397 \\
 V.Iyer and R.M.Wald, {\it Phys.Rev.} {\bf D50} (1994) 846
%
\bibitem{ref-gsl}
 J.D.Bekenstein, {\it Phys. Rev.} {\bf D7}(1973) 2333 \\
 J.D.Bekenstein, {\it Phys. Rev.} {\bf D9}(1974) 3292
%
\bibitem{ref-eit}
 D.Jou, J.Casas-V\'{a}zquez and G.Lebon, {\it Extended Irreversible Thermodynamics}, Springer, 1993
%
\bibitem{ref-sst}
 Y.Oono and M.Paniconi, {\it Prog.Theor.Phys.} {\bf suppl.130} (1998) 29 \\
 T.Hatano and S.Sasa, {\it Phys.Rev.Lett.} {\bf 86} (2001) 3463 \\
 S.Sasa and H.Tasaki, arXive cond-mat/0108365 \\
 S.Sasa and H.Tasaki, arXive cond-mat/0411052 \\
 K.Hayashi and S.Sasa, {\it Phys.Rev.} {\bf E68} (2003) 035104(R) \\
 K.Hayashi and S.Sasa, {\it Phys.Rev.} {\bf E69} (2004) 066119
%
\bibitem{ref-anti.sst}
 K.Heyon-Deuk and H.Hayakawa, {\it J.Phys.Soc.Jpn.} {\bf 72} (2003) 1904 \\
 K.Heyon-Deuk and H.Hayakawa, {\it J.Phys.Soc.Jpn.} {\bf 72} (2003) 2473 \\
 H.Hayakawa, T.H.Nishino and K.Heyon-Deuk, arXive cond-mat/0412011
%
\bibitem{ref-chandra}
 S.Chandrasekhar, {\it The Mathematical Theory of Black Holes}, Oxford Univ. Press, 1983
%
\bibitem{ref-sst.rad}
 H. Saida, {\it Physica} {\bf A356} (2005) 481
%
\bibitem{ref-gsl.sst}
 H. Saida, {\it Class.Quant.Grav.} {\bf 23} (2006) 6227
%
\bibitem{ref-evapo.sst}
 H. Saida, {\it Class.Quant.Grav.} {\bf 24} (2007) 691
%
\bibitem{ref-sgs}
 J.Binney and S.Tremaine, {\it Galactic Dynamics}, Princeton Univ. Press, 1987 \\
 R.Kippenhahn and A.Weigert, {\it Stellar Structure and Evolution}, Springer, 1994
%
\bibitem{ref-entropy}
 G.W.Gibbons and S.W.Hawking, {\it Phys.Rev.} {\bf D15} (1977) 2752
%
\bibitem{ref-noneq.rad}
 C.Essex, {\it Planet.Space.Sci.} {\bf 32} (1984) 1035 \\
 C.Essex, {\it Adv.Thermodynamics} {\bf 3} (1990) 435
%
\bibitem{ref-gsl.proof}
 W.G.Unruh and R.M.Wald, {\it Phys. Rev.}{ \bf D25}(1982) 942 \\
 W.G.Unruh and R.M.Wald, {\it Phys. Rev.} {\bf D27}(1982) 2271 \\
 E.E.Flanagan, D.Marolf and R.M.Wald, {\it Phys. Rev.} {\bf D62}(2000) 084035 \\
 V.P.Frolov and D.N.Page, {\it Phys.Rev.Lett.} {\bf 71}(1993) 3902
%
\bibitem{ref-info}
 S.W.Hawking, {\it Phys.Rev.} {\bf D14} (1976) 2460 \\
 U.H.Danielsson and M.Schiffer, {\it Phys.Rev.} {\bf D48} (1993) 4779 \\
 C.R.Stephens, G.'tHooft and B.F.Whiting, {\it Class.Quantum Grav.} {\bf 11} (1994) 621 \\
 A.Strominger, arXive hep-th/9501071 \\
 T.P.Shigh and C.Vaz, {\it Int.J.Mod.Phys.} {\bf D13} (2004) 2369
%
\end{thebibliography}
\end{document}